\documentclass[twocolumn,twoside,slac_two,superscriptaddress]{revtex4}
\usepackage{graphicx}
\usepackage{fancyhdr}
\usepackage{hyperref}
\hypersetup{
    colorlinks=true,
    urlcolor=magenta,
}
\usepackage{amsmath}
\newcommand{\galprop}{GALPROP}

\begin{document}

\title{HelMod: A Comprehensive Treatment of the Cosmic Ray Transport Through the Heliosphere}

\author{S. {Della~Torre}}
\email[corresponding author:]{stefano.dellatorre@mib.infn.it}
\affiliation{INFN Milano-Bicocca, Piazza della Scienza 3, 20125 Milano, Italy}
\author{M. Gervasi}
\affiliation{INFN Milano-Bicocca, Piazza della Scienza 3, 20125 Milano, Italy}
\affiliation{Physics Department, University of Milano-Bicocca, Milano, Italy}
\author{D. Grandi}
\affiliation{INFN Milano-Bicocca, Piazza della Scienza 3, 20125 Milano, Italy}
\author{G. Johannesson}
\affiliation{Science Institute, University of Iceland, Dunhaga 3, IS-107 Reykjavik, Iceland}
\author{G. La~Vacca} 
\affiliation{INFN Milano-Bicocca, Piazza della Scienza 3, 20125 Milano, Italy}
\author{N. Masi} 
\affiliation{INFN, Bologna, Italy}
\author{I.V. Moskalenko} 
\affiliation{Hansen Experimental Physics Laboratory, Stanford University, Stanford, CA 94305}
\affiliation{Kavli Institute for Particle Astrophysics and Cosmology, Stanford University, Stanford, CA 94305}
\author{E. Orlando}
\affiliation{Hansen Experimental Physics Laboratory, Stanford University, Stanford, CA 94305}
\affiliation{Kavli Institute for Particle Astrophysics and Cosmology, Stanford University, Stanford, CA 94305}
\author{T.A. Porter}
\affiliation{Hansen Experimental Physics Laboratory, Stanford University, Stanford, CA 94305}
\author{L. Quadrani}
\affiliation{INFN, Bologna, Italy}
\affiliation{Physics Department, University of Bologna, Bologna, Italy}
\author{P.G. Rancoita}
\affiliation{INFN Milano-Bicocca, Piazza della Scienza 3, 20125 Milano, Italy}
\author{D. Rozza}   
\affiliation{INFN Milano-Bicocca, Piazza della Scienza 3, 20125 Milano, Italy}
\affiliation{Physics Department, University of Milano-Bicocca, Milano, Italy}

%

\begin{abstract}
HelMod is a code evaluating the transport of Galactic cosmic rays through the inner heliosphere down to Earth. 
It is based on a 2-D Monte Carlo approach and includes a general description of the symmetric and antisymmetric 
parts of the diffusion tensor, thus, properly treating the particle drift effects. 
The model has been tuned in order to fit the data observed outside the ecliptic plane at several distances 
from the Earth and the spectra observed near the Earth for both, high and low solar activity levels.
A stand-alone python module, fully compatible with GalProp, was developed for a comprehensive calculation of solar modulation effects, 
resulting in a newly suggested set of local interstellar spectra.
\end{abstract}

\maketitle

\thispagestyle{fancy}


The intensity of Galactic Cosmic Rays (GCRs) observed at Earth vary with time accordingly to solar activity.
The overall effect of heliospheric propagation on the spectra of GCRs is called solar modulation.
Particle propagation in the heliosphere is affected by the outwards flowing solar wind (SW) 
with its embedded magnetic-field and magnetic-field irregularities. 
The so-generated and transported heliospheric magnetic field (HMF) is characterized by both 
the large scale structure (SW expansion from a rotating source) 
and low scale irregularities that vary with time according to the solar activity 
(e.g., by modification of SW velocity or local perturbations related to coronal mass ejection [CME]). 
The SW expansion causes CRs to propagate in a moving medium which is accounted for in the transport equation by the diffusion and adiabatic energy loss terms.
In addition, according to the original formulation by Ref.~\cite{parker58}, the HMF follows an Archimedean spiral that causes charged particles (i.e., CRs) to experience a combination of gradient, curvature and current sheet drifts~\cite{Jokipii77},
whose experimental evidence was provided, for instance, by Refs.~\cite{1986JGR....91.2858G,BoellaEtAl2001}.
The HelMod model was built ~\cite{Gervasi1999,1999NuPhS..78...26G,2003ESASP.535..637B,Bobik2004,Bobik2006,symposium2008,DellaTorre2009,AstraArticle2011,Bobik2011ApJ,DellaTorre2013AdvAstro,DellaTorre2016_OneD}, to include all individual processes, and, therefore, provides a realistic and unique description of the solar modulation. Here we provide a short description of the HelMod model\cite{helmodSite} (version 3.0), more details can be found in Refs.~\cite{Bobik2011ApJ,DellaTorre2013AdvAstro,Boschini2017AdvSR}.

\section{HelMod Model for GCR Propagation in the Heliosphere\label{Sect::Helmod}}
\begin{figure}[tb]
\centerline{
\includegraphics[width=0.48\textwidth]{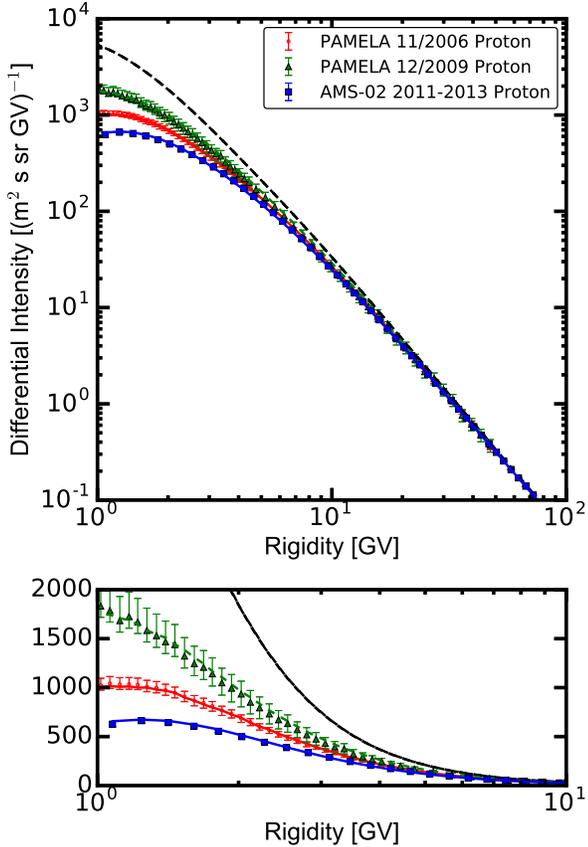}
}
\caption{Upper panel: Differential intensity of galactic proton  measured by PAMELA and AMS-02 compared with modulated spectra from
HelMod for the low and high activity respectively; the dashed lines are the GALPROP LIS’s (see text). The bottom panel show a zoom in linear scale  for rigidity range from 1 up to 10 GV.\label{fig:HelModMod}
}
\end{figure}
CR propagation in the heliosphere was first studied by Ref.~\cite{1965P&SS...13....9P}, who formulated the transport equation, also referred to as Parker Equation~(see, e.g., discussion in Ref.~\cite{Bobik2011ApJ} and reference therein):
\begin{align}
\label{EQ::FPE}
 \frac{\partial U}{\partial t}= &\frac{\partial}{\partial x_i} \left( K^S_{ij}\frac{\partial \mathrm{U} }{\partial x_j}\right)\\
&+\frac{1}{3}\frac{\partial V_{ \mathrm{sw},i} }{\partial x_i} \frac{\partial }{\partial T}\left(\alpha_{\mathrm{rel} }T\mathrm{U} \right)
- \frac{\partial}{\partial x_i} [ (V_{ \mathrm{sw},i}+v_{d,i})\mathrm{U}],\nonumber
\end{align}
where $U$ is the number density of Galactic particles per unit of kinetic energy $T$, $t$ is time, $V_{ \mathrm{sw},i}$ is the solar wind velocity along the axis $x_i$, $K^S_{ij}$ is the symmetric part of the diffusion tensor,
$v_{d,i}$ is the particle magnetic drift velocity (related to the antisymmetric part of the diffusion tensor),
and finally $\alpha_{\mathrm{rel} }=\frac{T+2m_r c^2}{T+m_r c^2} $, with $m_r$ particle rest mass in unit of
GeV/nucleon.

In the present work, for rigidity greater than 1\,GV, we use a functional form with a rigidity dependence following the one presented in Ref.~\cite{BurgerHattingh1998}:
\begin{equation}\label{EQ::KparActual}
 K_{||}=\frac{\beta}{3} K_0\left[ \frac{P}{1\text{GV}}+g_{\rm low}\right] \left(1+\frac{r}{\text{1 AU}}\right),
\end{equation}
where  $K_0$ is the diffusion parameter, which depends on the solar activity and magnetic polarity, $\beta$ is the particle speed in units of the speed of light, $P=qc/|Z|e$ is the particle rigidity expressed in GV, $r$ is the
heliocentric distance from the Sun in AU, and, finally, $g_{\rm low}$ is a parameter, which depends on the level of solar activity and allows the description of of the flattening with rigidity below few GV.
it is equal to 0.2 for low activity periods while for high activity periods $g_{\rm low}=0$ . The latter  is in qualitative agreement with simulations performed for
strong turbulence conditions as shown in Figures 3.5 and 6.5 of Ref.~\cite{shalchi2009}. 

The perpendicular diffusion coefficient is taken to be proportional to $K_{||}$ with a ratio
$K_{\perp,i}/K_{||}=\rho_i$ for both $r$ and $\theta$ $i$-coordinates (e.g., see Refs.~\cite{potgieter2000, BurgerHattingh1998} and references therein).
This description is consistent at high rigidity with those from quasi-linear theories (QLTs).
\citet{Palmer1982} constrains the value of $\rho_i$ between 0.02 and 0.08 at Earth.
We found  best agreement at $\rho_i\approx 0.058$.
As remarked in Ref.~\cite{DellaTorre2013AdvAstro}, in this description $K_{||}$
has no latitudinal dependence and a radial dependence $\propto r$;
nevertheless, the reference frame transformation between the field aligned to the spherical
heliocentric frame (see, e.g., Ref.~\cite{burg2008}) introduces a polar angle dependence.


As discussed in Section 2.1 of Ref.~\cite{Bobik2011ApJ}, the diffusion parameter, $K_0$, provides a scaling factor for the overall modulation intensity and is evaluated 
using the Monthly Smoothed Sunspot Number. Therefore,
the effective modulation experienced by CRs is related to the solar activity and polarity of the magnetic field.
This approximation is valid as long as disturbances coming from the Sun
(like CMEs) are short and not very frequent, and
do not significantly affect the average behavior of the heliospheric medium.

During periods of high solar activity the rate of CMEs increases
leading to a more chaotic structure of magnetic field and stronger turbulence, thus the HMF cannot be properly described by a dipole configuration.
To improve the practical relationship between $K_0$ and solar activity, we use
the Neutron Monitor Counting Rate (NMCR). In the current work, we exploit the NMCR recorded by the McMurdo station and available through the Neutron Monitor Database~\cite{nmdbWeb} following the same fitting procedure used in Section 2.1 of Ref.~\cite{Bobik2011ApJ}. NMCR allows us to account for short-time and large-scale variations occurring during the high solar activity periods, and thus to re-scale the diffusion parameter accordingly.

In the present work, we use the drift model originally developed by Ref.~\cite{Potgieter85}
and refined using definitions of Parker's magnetic field with polar correction
as reported in Ref.~\cite{DellaTorre2013AdvAstro} (see also Ref.~\cite{Raath2016} for a discussion about modified Parker's magnetic field).
Since during the high activity period the HMF is far from
being considered regular, in this work we introduced a correction factor that
suppresses any drift velocity at solar maximum.

We compute the CR propagation from the Termination Shock (TS)
down to Earth using a Monte Carlo approach, i.e.,
the HelMod Monte Carlo code~\cite{Bobik2011ApJ} that solves the
two-dimensional Parker equation for CR transport through the heliosphere.
HelMod code applies the stochastic integration to a set of stochastic differential equations (SDEs),
which are fully equivalent to Eq.~(\ref{EQ::FPE}) (see a discussion in, e.g., Refs.~\cite{Bobik2011ApJ,DellaTorre2016_OneD}).
In this scheme, quasi-particle objects evolve \emph{backward-in-time}
from the location of the detector, i.e., from the Earth, back to the TS.
The modulated spectrum is then obtained by averaging the evaluated LIS fluxes, which take into account the reconstructed rigidity at the heliospheric
boundary (see Section 4.1.2 in Ref.~\cite{DellaTorre2016_OneD}).


In the current calculation, we assume a static and spherical heliosphere with TS located at 100 AU \cite{Bobik2011ApJ}.
Even though variations of the real size of the heliosphere may be important for the analysis of CR propagation near the TS, we do not consider them in this work.

The numerical uncertainties of our Monte Carlo approach were evaluated in Ref.~\cite{DellaTorre2016_OneD}, who employed the Crank-Nicholson technique for the SDE integration, and found them to be less than 0.5\% at low rigidities. The large number of simulated events ensure that the statistical errors are negligible compared to the systematic uncertainties.
HelMod parameters were tuned using proton data in both high and low solar activity (see Fig.~\ref{fig:HelModMod}). 
The local interstellar spectrum (LIS) is assumed to be isotropic along the heliosphere boundary. 
Nowadays LIS parametrization is constrained  by measurements from Voyager probes, at low energy, and AMS-02, at high energy. 
In Ref.~\cite{MasiECRS2016} it is discussed how further constrains can be achieved requiring the agreement of \galprop{} LIS, modulated using HelMod, with a large set experimental data in different solar conditions.


\section{HelMod Python Module for \galprop{}}\label{Sect::PythonM}

\begin{figure}[tb]
\centerline{
\includegraphics[width=0.5\textwidth]{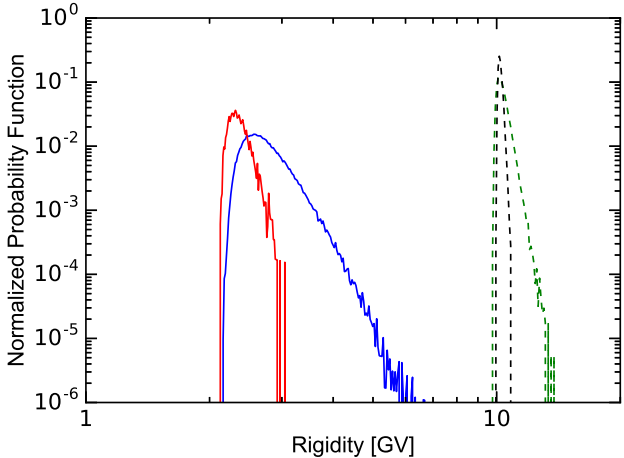}
}
\caption{The computed normalized probability function $G(P_0|P)$ for $P_0=2$ (solid lines) and $9.7$ GV (dashed lines) evaluated for AMS-02 proton binning during the period 2011--2014 (blue and green) and for PAMELA proton binning during December 2009 (red and black), see text for details.
}
\label{fig:PythonModule_G}
\end{figure}
The SDE integration with HelMod results in quite an expensive effort from the computational point of view since minimization of uncertainties requires a simulation of a considerable number of events propagating from the Earth to the heliospheric boundary.
The common approach is to evaluate the modulated spectrum directly from the numerical integration using the procedure described in Ref.~\cite{DellaTorre2016_OneD} (and references therein) that forces a new simulation run for each LIS to be tested.
A different approach allows to use SDE Monte Carlo integration to evaluate the normalized probability
function $G(P_0|P)$ that gives a probability for a particle observed at the Earth with a rigidity $P_0$ having a rigidity $P$ at the heliospheric boundary.
Once $G(P_0|P)$ is evaluated it is possible to obtain the modulated spectrum directly from $J_{\rm LIS}$ provided by \galprop{} with the relationship reported in Ref.~\cite{PeiBurger2010}.
\begin{equation}\label{eq::PyMod_modulation}
 J_{\rm mod}(P_0)= \int_0^\infty J_{\rm LIS}(P)G(P_0|P)dP
\end{equation}

For illustration, in Figure~\ref{fig:PythonModule_G} we show the computed
normalized probability function for $P_0=2$ and $9.7$ GV evaluated for protons
during the period 2011-2014, equivalent to the data taking period of released AMS-02 data~\cite{2015PhRvL.114q1103A}.
The normalized probability functions were evaluated for several CR species (p, He, B, C, e$^-$) for a set of selected experiments running from 1997 to 2015~\cite{bess_prot,AMS01_prot,PamelaProt2013,2014A&A...569A..32M,2015PhRvL.114q1103A,BESS2007_Abe_2016}.

To simplify the calculations,
we developed a python script that reads the \galprop{} output and provides the modulated spectrum for periods of selected experiments. The calculation of propagation in the heliosphere is substituted by the integration of Eq.~(\ref{eq::PyMod_modulation}) with the normalized probability functions, which are pre-evaluated using the HelMod code as described in the previous Section. This method dramatically accelerates the modulation calculations while provides the accuracy of the full-scale simulation.

The HelMod python module can be downloaded from a dedicated website or used on-line~\cite{helmodSite}. It reads the \galprop{} output format (FITS) and provides a modulated spectrum for a specified period of time. While the heliospheric propagation is fixed by using the provided functions $G(P_0|P)$, the LIS spectrum can be specified by a user. The output rigidity binning is chosen to be compared with CR experiments in the specified period, alternatively the AMS-02 rigidity binning is chosen as the standard one.

\section{Conclusion}
We presented the 2-D Monte Carlo Heliospheric Modulation Model, i.e. HelMod, to evaluate GCR modulated spectra at Earth.
The model includes details of individual processes occurring in the heliosphere particle propagation, thus providing a realistic and unique description of the solar modulation. The model is tuned by using several proton observations at Earth in different solar activity periods, treating in a unique description the high and low solar activity periods. Moreover HelMod provides further constraints to proton LIS in conjunction with \galprop.
To make the calculation available to the community, a python module and a dedicated website, are developed to provide modulated spectra for selected GCR experiments in different solar activity periods.

\bigskip 
\begin{acknowledgments}
We wish to specially thank Pavol Bobik, Giuliano Boella, Matteo Boschini, Karel Kudela, Simonetta Pensotti, Marian Putis, Mauro Tacconi and Mario Zannoni for their support to the HelMod code and many useful suggestions.
This work is supported by ASI (Agenzia Spaziale Italiana) under contract ASI-INFN I/002/13/0 and ESA (European Space Agency) contract 4000116146/16/NL/HK. 
 Igor Moskalenko, Elena Orlando, Troy Porter acknowledge support from NASA Grants Nos.~NNX13AC47G and NNX17AB48G, 
 Elena Orlando additionally acknowledges support from NASA Grants Nos.~NNX16AF27G and NNX15AU79G, 
 and Troy Porter additionally acknowledges support from NASA Grant No.~NNX10AE78G. 
\end{acknowledgments}

\bigskip 
\bibliography{bibliography}
\end{document}